\documentclass[12pt, epsfig]{article}
\usepackage{amsmath,amsfonts,amssymb,amsthm,amstext,amscd,eucal,graphicx}
\usepackage[all]{xy}
\usepackage{dsfont}
\usepackage{hyperref}
\usepackage{amsmath}
\usepackage{slashed}
\usepackage{color}
\usepackage{epsfig}%
\usepackage[active]{srcltx}
\usepackage{enumitem}
\setlist{nolistsep}

\makeatletter \@addtoreset{equation}{section}

\makeatletter\renewcommand\section{\@startsection {section}{1}{\z@}%
                                   {-3.5ex \@plus -1ex \@minus -.2ex}
                                   {2.3ex \@plus.2ex}%
                                   {\normalfont\large\bfseries}}
\renewcommand\subsection{\@startsection{subsection}{2}{\z@}%
                                     {-3.25ex\@plus -1ex \@minus -.2ex}%
                                     {1.5ex \@plus .2ex}%
                                     {\normalfont\bfseries}}

\newcommand{\be}{\begin{equation}}
\newcommand{\ee}{\end{equation}}
\newcommand{\bea}{\begin{eqnarray}}
\newcommand{\eea}{\end{eqnarray}}
\newcommand{\bse}{\begin{subequations}}
\newcommand{\ese}{\end{subequations}}
\newcommand{\beqa}{\begin{eqnarray}}
\newcommand{\eeqa}{\end{eqnarray}}
\newcommand{\beqar}{\begin{eqnarray*}}
\newcommand{\eeqar}{\end{eqnarray*}}
\newcommand{\bi}{\begin{itemize}}
\newcommand{\ei}{\end{itemize}}
\newcommand{\bn}{\begin{enumerate}}
\newcommand{\en}{\end{enumerate}}

\newcommand{\ba}{\begin{array}}
\newcommand{\ea}{\end{array}}
\newcommand{\bc}{\begin{center}}
\newcommand{\ec}{\end{center}}

\newcommand{\nn}{\nonumber}

\def\dd{\textrm{d}}
\def\nn{\nonumber \\}

\definecolor{darkgreen}{rgb}{0,0.3,0}
\definecolor{darkblue}{rgb}{0,0,0.3}
\definecolor{darkred}{rgb}{0.7,0,0}

\begin{document}
\begin{titlepage}

\vfill
\begin{flushright}
DMUS--MP--16/20 \\
APCTP Pre2016-019 
\end{flushright}

\vfill

\begin{center}
   \baselineskip=16pt
   {\Large {\bf  Ba\~nados and SUSY: }}
  \vskip 2mm
   { \large {\bf On Supersymmetry and Minimal Surfaces of \\ Locally AdS$_3$ Geometries}}
 \vskip 2cm
     E. \'O Colg\'ain$^{a, b}$, H. Yavartanoo$^{c}$

       \vskip .6cm
             \begin{small}
             \textit{$^a$ Department of Mathematics, University of Surrey, Guildford GU2 7XH, UK}
                 \vspace{3mm}
                 
             \textit{$^b$ Asia Pacific Center for Theoretical Physics \& \\ Department of Physics, POSTECH, Pohang 37673, Korea}
                 \vspace{3mm}

   \textit{$^c$ \it  State Key Laboratory of Theoretical Physics, Institute of Theoretical Physics, \\
Chinese Academy of Sciences, Beijing 100190, China.
 }
                 \vspace{3mm}

             \end{small}
\end{center}

\vfill \begin{center} \textbf{Abstract}\end{center} \begin{quote}
We extend the classification of supersymmetric locally AdS$_3$ geometries, beyond BTZ black holes, to the Ba\~nados geometries, noting that supersymmetries are in one-to-one correspondence with solutions to the Hill differential condition. We show that the number of global supersymmetries is an orbit invariant quantity and identify geometries with zero, one, two, three and four global supersymmetries. As an application of our classification, we exploit supersymmetry, which is preserved locally in the bulk, to determine space-like co-dimension two surfaces in AdS$_3$. In the process, we by-pass geodesics or mappings of AdS$_3$, neither of which have an analogue in higher dimensions, to recover known Hubeny-Rangamani-Takayanagi surfaces. Our findings suggest supersymmetry may be exploited to find extremal surfaces in holographic entanglement entropy. \end{quote}

\end{titlepage}


\section{Introduction}
The Ba\~nados geometries \cite{Banados:1998gg} encompass the most general locally AdS$_3$ solutions to Einstein gravity with a negative cosmological constant  that satisfy Brown-Henneaux boundary conditions \cite{Brown:1986nw}. The spacetime is characterised by two holomorphic periodic functions, $L_{\pm}$, where the simplest examples include BTZ black holes \cite{Banados:1992wn, Banados:1992gq} with positive constant $L_{\pm}$. The geometries exhibit locally an $SL(2, \mathbb{R}) \times SL(2,\mathbb{R})$ symmetry, which is broken to the Cartan $U(1) \times U(1)$ by global boundary conditions. 

Just as local symmetries are broken globally, the Ba\~nados geometries may also be embedded in 3D $\mathcal{N} = (1,1)$ AdS$_3$ supergravity \cite{Achucarro:1987vz}, where again, the preservation of supersymmetry depends on the effect of the boundary conditions on the Killing spinors. For BTZ black holes, global supersymmetry has been classified \cite{Coussaert:1993jp}, resulting in the conclusion that massless, extremal and generic BTZ preserve globally two, one and zero supersymmetries, respectively. 

In this work, we extend the classification of globally supersymmetric geometries to the larger class of Ba\~nados geometries. In general, we show that local solutions to the Killing spinor equation are in one-to-one correspondence with solutions to the Hill differential equation (see \cite{Hill} for a review). As a consistency check on the Killing spinors, we demonstrate that one can recover the local isometries directly from vector bi-linears. Examining the global properties of the spinors, we provide a complete classification of supersymmetric Ba\~nados geometries. As we will argue, all geometries in the same Virasoro orbit \cite{Balog:1997zz, Sheikh-Jabbari:2016unm}, carry the same amount of supersymmetry. In other words, the number of global supersymmetries is an orbit invariant quantity. Therefore, by analysing Killing spinors in geometries corresponding to representatives of Virasoro co-adjoint orbits, we get the complete classification of supersymmetric Ba\~nados geometries preserving one, two, three and four supercharges. 
 
One take-home message of the classification is that supersymmetry is always present at a local level, but it may be broken globally. However, it is fitting to contemplate an interesting application for this local supersymmetry, and one setting where it promises to be useful is the determination of co-dimension two surfaces in computing entanglement in holographic theories. To justify this assertion, we note that supersymmetric surfaces in Riemannian manifolds correspond to volume minimising surfaces, or \textit{calibrated cycles} \cite{Harvey:1982xk} (also \cite{Gauntlett:2003di}). 

We recall that Ryu-Takayanagi (RT) prescription \cite{Ryu:2006bv, Ryu:2006ef} provides a simple way to determine the entanglement entropy of a CFT, which is generally tricky beyond 2D \cite{Holzhey:1994we, Calabrese:2004eu, Calabrese:2009qy}, by mapping the problem to a calculation of the area of a minimal surface in the AdS bulk spacetime. The RT conjecture for AdS$_3$ gravity was initially proved in \cite{Faulkner:2013yia, Hartman:2013mia}, and an explanation in terms of AdS/CFT in general dimensions appeared later in \cite{Lewkowycz:2013nqa}. One limitiation of the RT proposal is that it does not hold for time-dependent states and a covariant generalisation was proposed by Hubeny-Rangamani-Takayanagi (HRT) \cite{Hubeny:2007xt}  \footnote{See \cite{Dong:2016hjy} for a recent gravitational argument supporting the HRT proposal. }. Unfortunately, taking the HRT proposal at face value, it is technically challenging to determine extremal surfaces analytically, since the equations are second order. 

Thankfully, life has been kind and theoretical physicists have developed supersymmetry \footnote{Admittedly, with loftier goals in mind.}.  One spin-off of supersymmetry is precisely that second order equations can be replaced by first order equations, and it is surprising that supersymmetric surfaces have not been considered before in the context of holographic entanglement entropy. Elsewhere, supersymmetry has been extremely successful in identifying geometries, with the discovery of the Sasaki-Einstein metrics $Y^{p,q}$ \cite{Gauntlett:2004zh, Gauntlett:2004yd} providing a striking example. It is difficult to believe that this class of Sasaki-Einstein spaces would have been found by solving second order equations. So, between static spacetimes, where the RT proposal is simple and works well, and a generic time-dependent spacetime, which is messy, but in principle the HRT proposal solves the problem, one may expect to find a Goldilocks zone of time-dependent supersymmetric spacetimes, where supersymmetry serves to simplify the identification of extremal surfaces. 

In this work, we make an initial foray in this direction using AdS$_3$ as a guinea pig for our ideas. In this unique setting, we know that the co-dimension two surfaces we seek are simply space-like geodesics. However, with an eye to higher-dimensional generalisations, we will eschew solving the geodesic equation in favour of extracting the information directly from supersymmetry. As mentioned earlier, this is expected to work for minimal surfaces in static spacetimes, yet for extremal surfaces in time-dependent spacetimes, we anticipate our results are novel. 

In practice, we introduce a projection condition, which generically preserves two supersymmetries locally, and show that it recovers known RT embeddings for both massless and static BTZ. For the rotating BTZ black hole, we remark that there is no mathematical analogue of calibrated cycles for pseudo-Riemannian manifolds, yet our projection condition recovers, and generalises the HRT embedding \footnote{We recover space-like geodesics allowing two independent constants of motion under the assumption that the velocity vector along the curve has unit norm.}. Moreover, with the assumption of constant $\tau$ (where $\tau$ is a time-like direction defined in the Ban\~ados geometries \cite{Sheikh-Jabbari:2016unm}), we  identify the extremal surfaces for the general Ba\~nados class. We check that the supersymmetric embeddings agree with space-like geodesics, but we emphasise that we have neither assumed that the Ba\~nados geometries are locally equivalent to AdS$_3$ in Poincar\'e patch, nor have we relied on geodesics. We remark that proper local coordinate transformations have already been exploited to obtain minimal surface for the general Ba\~nados geometry \cite{Sheikh-Jabbari:2016znt} \footnote{See \cite{Roberts:2012aq, Maxfield:2014kra} for earlier work in this direction.}. As a result, we are led to believe that supersymmetry may be applied to higher-dimensional geometries, both AdS and non-AdS \cite{Anninos:2013nja, Castro:2015csg, Song:2016gtd}, to determine entanglement entropy holographically. 

The structure of this paper is as follows. In section \ref{sec:Hill} we introduce the Ba\~nados geometries in a supersymmetric context, where we solve the Killing spinors and reconstruct the AdS$_3$ isometries. In section \ref{sec:classification}, we extend the classification of supersymmetric solutions from BTZ black holes to the larger class of Ba\~nados geometries. In section \ref{sec:codim2}, we introduce a projection condition that isolates space-like co-dimension two surfaces in the bulk and show that it recovers both static and time-dependent geodesics, which correspond to extremal surfaces one should determine in the HRT proposal. Finally, we summarise our results and discuss future prospects. In the appendix, we collect space-like geodesics for BTZ black holes. 

\section{Hill equation \& supersymmetric AdS$_3$ geometries}
\label{sec:Hill}
To begin our story, we will show that a large class of (locally) supersymmetric AdS$_3$ geometries are in one-to-one correspondence with solutions to the Hill differential equation. We will use this in the next section to classify the geometries according to the preserved supersymmetry. We note that BTZ black holes were analysed in ref. \cite{Coussaert:1993jp} and our work constitutes an extension to the larger Ba\~{n}ados class of geometries.

Our point of departure is 3D $\mathcal{N} = (1,1)$ AdS$_3$ supergravity \cite{Achucarro:1987vz}, which is the simplest supersymmetric extension of Einstein gravity with a negative cosmological constant. The vanishing of the gravitini variation results in the usual Killing spinor equation for AdS$_3$, 
\be
\label{KSE_ads}
\nabla_{\mu} \epsilon_{\pm} = \pm \frac{1}{2 \ell} \gamma_{\mu} \epsilon_{\pm}, 
\ee
where we have retained the AdS$_3$ radius $\ell$, and the Killing spinors $\epsilon_{\pm}$ quantify the degree to which supersymmetry is preserved by a given spacetime geometry. In this context maximal supersymmetry corresponds to four supersymmetries, two for each Killing spinor. 

Using the Killing spinor equation, it is possible to show that the following vectors
\be
\label{vec_bilinear}
\mathcal{K}^{\pm}_{\mu} = \frac{i}{2} \bar{\epsilon}_{\pm} \gamma_{\mu} \epsilon_{\pm}, 
\ee
constructed from the spinors, satisfy the Killing equation and consequently generate isometries \footnote{Using Fierz identity and Killing spinor equation one can show that $\mathcal{K}^{\pm}$ are either time-like or null.}. As AdS$_3$ has $SL(2, \mathbb{R}) \times SL(2, \mathbb{R})$ isometry, it is necessary to entertain both signs above in (\ref{KSE_ads}) in order to realise all isometries from the Killing spinors \footnote{For supersymmetric AdS$_3$ geometries in higher dimensions, one can try to extract the corresponding superalgebra this way, but one typically only finds linear combinations of the Killing vectors \cite{Gauntlett:1998kc, Lozano:2015bra}.}. Like the Killing spinors, the vectors are only defined locally and whether they are globally defined depends on the properties of the spacetime. We will return to this point after we introduce the Ba\~{n}ados geometries. We remark that the two Killing spinor equations are related through an overall change in sign in the gamma matrices. 

Our task in this section is to solve the Killing spinor equation in the Ba\~nados geometries, which we now introduce. The most general AdS$_3$ solution to Einstein gravity with Brown-Henneaux boundary conditions may be expressed as 
\be
\label{general3dmetric}
\dd s^2= \ell^2\frac{ \dd \rho^2}{\rho^2}-(\rho \,  \dd x^+- \frac{\ell^2}{\rho} L_- \dd x^- )(\rho \, \dd x^- - \frac{ \ell^2}{\rho}L_+ \dd x^+ ), 
\ee
where $x^{\pm} \in [0, 2 \pi]$ and $L_+ \equiv L_+(x^+)$ and $L_- \equiv L_-(x^-)$ are arbitrary periodic functions. We will henceforth refer to these solutions as Ba\~nados geometries. 

In order to familiarise ourselves with these solutions, it is worth noting that $L_{\pm} = 0$ simply corresponds to the AdS$_3$ spacetime metric in Poincar\'e coordinates. Taking into account the periodicity of the coordinates $x^{\pm}$, the geometry is better referred to as massless BTZ. To recover the much-loved form of the rotating BTZ black hole, one takes $L_{\pm}$ to be positive constants. It is straightforward to check that the ADM form of the metric 
\be
\label{BTZ}
\dd s^2=-\frac{(r^2-r_+^2)(r^2-r_-^2)}{\ell^2 \, r^2} \dd t^2 +\frac{\ell^2 \, r^2 \dd r^2}{(r^2-r_+^2)(r^2-r_-^2)}+r^2\left({\dd x} + \frac{r_+ r_-}{\ell \, r^2} \dd t \right)^2,
\ee
may be found through the following coordinate transformations: 
\bea
\label{coordinatetrans}
\rho^2 &=& \frac{1}{4} \left( 2 r^2 - r_+^2 - r_-^2 \pm 2 \sqrt{(r^2 - r_+^2)(r^2-r_-^2)} \right), \nn
x^{\pm} &=& \frac{t}{\ell} \pm {x}, \quad L_{\pm} = \frac{1}{4 \ell^2} ( r_+ \pm r_-)^2. 
\eea
In the metric \eqref{BTZ},  the range of the $r$ coordinate is $[0,\infty)$, but to get real values in the coordinate transformation \eqref{coordinatetrans},  we must restrict to $r\geq r_+$. As a consequence, this gives the range $\rho\in [\rho_0, \infty)$, where $\rho_0=\ell (L_+L_-)^\frac{1}{4}$. In other words, for BTZ black hole, the coordinate system \eqref{general3dmetric} covers only the region outside the outer horizon. It is also clear from analytical continuation in $\rho$ that one can cover the region inside the inner horizon as well, however this does not cover the region between the two horizons. A similar argument can be applied for general Ba\~nados solutions \cite{Sheikh-Jabbari:2016unm}.

If one demands extremality, it is clear from (\ref{coordinatetrans}) that this corresponds to setting one of $L_{\pm}$ to zero. 

As stated earlier, for generic functions $L_{\pm} (x^{\pm})$, the Ba\~{n}ados geometries admit six local Killing vector fields, which correspond to the $SL(2, \mathbb{R}) \times SL(2, \mathbb{R})$ symmetry. These six Killing vectors may be expressed as \cite{Sheikh-Jabbari:2016unm, Sheikh-Jabbari:2014nya} 
\bea
\chi[K_+, K_-] = \chi^{\rho} \partial_{\rho} + \chi^+ \partial_+ + \chi^- \partial_-, 
\eea
where we have defined: 
\be
\label{Killing_vec}
\chi^{\rho} = - \frac{\rho}{2} ( K'_++K'_-), \quad \chi^{\pm} = K_{\pm} + \frac{ \ell^2 \rho^2 K''_{\mp} + \ell^4 L_{\mp} K''_{\pm}}{2 (\rho^4 - \ell^4 L_+ L_-)}, 
\ee
in terms of functions $K_{\pm}(x^{\pm})$, and primes denote derivatives with respect to the argument of $K_{\pm}$. An analysis of the Killing equation reveals that $K_{\pm}$ obey the third order differential equation \cite{Sheikh-Jabbari:2016unm, Sheikh-Jabbari:2014nya}, 
\be
\label{3rd_order_eqn}
K'''_{\pm} - 4 K'_{\pm} L_{\pm} - 2 K_{\pm} L'_{\pm} = 0. 
\ee
We emphasise that the concept of a Killing vector is purely a local one and every solution to (\ref{3rd_order_eqn}) generates a local Killing vector. The above third order equation has six linearly independent solutions, so we can be confident of recovering all the isometries of AdS$_3$. However, due to periodicity constraints, not all the Killing vectors are globally defined. 

Admittedly, third order equations can be a little off-putting. In fact, it can be shown that solutions to the above equation can be specified by solutions to the following Hill equations: 
\be
\label{Hill} 
\psi''_{\pm} = L_{\pm} \psi_{\pm}. 
\ee
To be more precise, since the Hill equation is second order, we can introduce two linearly independent solutions to the Hill equation, denoted $\psi^i_{\pm}, ~i = 1, 2$. From these solutions, we construct three solutions to each equation (\ref{3rd_order_eqn}), 
\be
\label{vector_functions}
K^0_{\pm} = \psi^1_{\pm} \psi^2_{\pm}, \quad K^1_{\pm} = \psi^1_{\pm} \psi^1_{\pm}, \quad K^{2}_{\pm} = \psi^2_{\pm} \psi^2_{\pm}, 
\ee
resulting in six Killing directions, as advertised. However, recalling the Floquet theorem \cite{Hill} (see appendix A), one finds that only two Killing vectors are globally defined \cite{Sheikh-Jabbari:2016unm, Sheikh-Jabbari:2014nya}.

\subsection{Solving Killing spinor equation}

Having introduced the Ba\~{n}ados geometries (\ref{general3dmetric}), we proceed to solve the Killing spinor equation (\ref{KSE_ads}). To be concrete, we introduce the following gamma matrices, 
\be
\label{gamma_matrices}
 \gamma_+ =\begin{pmatrix}
   0  &  \sqrt{2} i  \\
   0   &  0
\end{pmatrix},\quad  \gamma_- =\begin{pmatrix}
   0   &  0  \\
  \sqrt{2}  i  &  0
\end{pmatrix},
\quad 
\gamma_r=\begin{pmatrix}
   1   &  0  \\
   0   &  -1
\end{pmatrix}, 
\quad
\ee
which form a representation of the Clifford algebra, $\{ \gamma_{\mu}, \gamma_{\nu} \} = 2 \eta_{\mu \nu}$, where 
\be
 \eta_{IJ}=\begin{pmatrix}
   1   &  0 &0 \\
   0   &  0&-1 \\
   0 & -1 &0
\end{pmatrix}. 
\ee
In terms of a representation of Cliff(2,1), i.e. $\eta_{\mu \nu} = \textrm{diag} (-1, 1, 1)$, the gamma matrices can be rewritten as  
\be
\gamma_0 = \frac{1}{\sqrt{2}} (\gamma_+ + \gamma_-), \quad \gamma_1 = \frac{1}{\sqrt{2}} (\gamma_+ - \gamma_-), 
\ee
so that $\gamma_{0} = i \sigma_1$, $\gamma_1 = - \sigma_2$ and $\gamma_r = \sigma_3$ in terms of Pauli matrices. The product of the gamma matrices become $\gamma_{01r} = +1$. To solve the Killing spinor equation, we will use the same gamma matrices throughout, but will simply flip the sign in the AdS$_3$ Killing spinor equation (\ref{KSE_ads}). Upon the introduction of the natural dreibein 
\be
e^{\rho}= \ell \frac{\dd \rho}{\rho},\quad e^+=\frac{1}{\sqrt2}\left( \rho \, \dd x^+-\frac{\ell^2}{\rho} L_- \dd x^-\right),\quad e^- =\frac{1}{\sqrt2}\left( \rho \, \dd x^- -\frac{\ell^2}{\rho} L_+ \dd x^+\right), 
\ee
a straightforward calculation allows us to solve for the Killing spinors, which take the form 
\be
\epsilon_{+} = \rho^{ \frac{1}{2}} \eta_+ + \rho^{- \frac{1}{2}} \eta_-, \quad \epsilon_{-} = \rho^{ -\frac{1}{2}} \tilde{\eta}_+ + \rho^{+ \frac{1}{2}} \tilde{\eta}_-, 
\ee
where $\eta_{\pm}$ and $\tilde{\eta}_{\pm}$ are chiral spinors, i. e. $\gamma_{\rho} \eta_{\pm} = \pm \eta_{\pm}$, $\gamma_{\rho} \tilde{\eta}_{\pm} = \pm \tilde{\eta}_{\pm}$, which \textit{a priori} depend on $x^{\pm}$. Solving for the dependence on $x^{\pm}$, we find the following equations: 
\bea
\partial_- \eta_{\pm} &=& 0, \quad \sqrt{2} \partial_{+} \eta_+ = \frac{1}{\ell} \gamma_{+} \eta_-, \quad \sqrt{2} \partial_{+} \eta_- = - \ell L_+ \gamma_- \eta_+, \\
\partial_+ \tilde{\eta}_{\pm} &=& 0, \quad  \sqrt{2} \partial_{-} \tilde{\eta}_- = -\frac{1}{\ell} \gamma_{-} \tilde{\eta}_+, \quad \sqrt{2} \partial_{-} \tilde{\eta}_+ = \ell L_- \gamma_+ \tilde{\eta}_-. 
\eea
It is worth noting that the $\epsilon_+$ Killing spinor only depends on $x^+$ and the $\epsilon_-$ spinor only depends on $x^-$. As a result, it is obvious that we need to consider both signs in the Killing spinor equation in order to reconstruct all isometries. Combining the above equations, we find two copies of the Hill equation (\ref{Hill}), namely 
\be
\partial_{+}^2 \eta_+ = L_+ \eta_+, \quad \partial_-^2 \tilde{\eta}_- = L_- \tilde{\eta}_-. 
\ee
It is worth noting that by invoking supersymmetry, which like the Killing vectors is locally defined, we can circumvent the unsightly third order equation (\ref{3rd_order_eqn}). In contrast, we have found that two copies of the Hill equation - each one corresponding to an $SL(2, \mathbb{R})$ symmetry - naturally emerge from the Killing spinor equation. For consistency, we should check that the vector bi-linears (\ref{vec_bilinear}) agree with the Killing vectors (\ref{Killing_vec}).

But, before doing so we quickly summarise the solution to the Killing spinor equation in the Ba\~nados geometries. Introducing the explicit form for the remaining gamma matrices, the final solution to the Killing spinor equation (\ref{KSE_ads}) may be written as 
\bea
\label{KS}
\epsilon^i_+ = \left( \begin{array}{c} \rho^{\frac{1}{2}} \, \psi^i_+ \\ - i \ell \rho^{-\frac{1}{2}} \, \partial_+ \psi^i_+ \end{array} \right), \quad \epsilon_-^i = \left( \begin{array}{c}  i \ell \rho^{-\frac{1}{2}} \, \partial_- \psi^i_- \\ \rho^{\frac{1}{2}} \, \psi^i_- \end{array} \right), 
\eea
where we have added a superscript to label the two linearly independent solutions to the Hill equation. 

\subsection{Killing vectors}
For consistency, we will demonstrate that the Killing vectors can be reconstructed from the spinors. To be concrete, we focus on the vector bi-linear constructed from $\epsilon^1_{+}$ and $\epsilon^2_{+}$ (\ref{KS}). Taking other combinations of the spinors in turn, one can recover the full set of Killing vectors, but the analysis is repetitive and we omit the details. 

Without loss of generality, we will assume that $\psi^i_{+}$ are real\footnote{Note that even for real functions $L_\pm$, the Floquet index generically is a complex number. Therefore, if we choose real solutions to the Hill equation, they may not follow the Floquet theorem. However, if we take the Floquet form of $\psi_{\pm}$,  the functions $\mathcal{K}^0_{\pm}$ are still real and periodic, and the corresponding Killing vectors are real and periodic as a result.}, since confronted with complex solutions to the Hill equation, we always have the freedom to extract the real and imaginary parts. Our choice of gamma matrices (\ref{gamma_matrices}), allows us to define $\bar{\epsilon} = \epsilon^{\dagger} \sigma_1$, with $A = \sigma_1$ playing the role of the inter-twiner relating different representations of the Clifford algebra. As a further consequence of our gamma matrices, their symmetry properties require us to define our Killing vectors  with an overall $i$ factor to ensure that they are real,   
\be
(\mathcal{K}^0_{+})^{\mu} = \frac{i}{2} \bar{\epsilon}^1_{+} \gamma^{\mu} \epsilon^2_+. 
\ee
The factor of two is added to facilitate comparison. 

Note that our labelling here intentionally mirrors the earlier functions $K_{\pm}$ appearing in the Killing vector (\ref{Killing_vec}), with the slight difference that here we are denoting the vector and not the functions in the vector. Substituting for the gamma matrices, we can evaluate the components of the vector: 
\bea
(\mathcal{K}^0_{+})^{\rho} &=& - \frac{\ell}{2} ( \psi^1_{+} \psi^2_{+} )', \quad (\mathcal{K}^0_{+})^+ =  \frac{\rho}{\sqrt{2}} \psi_+^1 \psi^2_+, \quad  (\mathcal{K}^0_{+})^- = \frac{\ell^2}{ \sqrt{2} \rho} (\psi_+^1)' (\psi_+^2)'. 
\eea
Next, we can introduce the inverse dreibein, 
\be
\bar{\partial}_{\rho} = \frac{\rho}{\ell} \partial_{\rho}, \quad 
\bar{\partial}_{\pm} = \frac{\sqrt{2} \rho^2}{(\rho^4 - \ell^4 L_+ L_-)} \left( \rho \partial_{\pm} + \frac{\ell^2 L_{\pm}}{\rho} \partial_{\mp} \right), 
\ee
to check that the vector takes the expected form 
\bea\label{PKV}
\mathcal{K}^0_+ &=& - \frac{\ell}{2} ( \psi^1_{+} \psi^2_{+} )' \, \bar{\partial}_{\rho} + \frac{\rho}{\sqrt{2}} \psi_+^1 \psi^2_+ \, \bar{\partial}_+ + \frac{\ell^2}{ \sqrt{2} \rho} (\psi_+^1)' (\psi_+^2)' \, \bar{\partial}_-, \nn
&=& - \frac{\rho}{2} ( \psi^1_{+} \psi^2_{+} )' \, \partial_{\rho} + \left( \psi_+^1 \psi_+^2 + \frac{\ell^4 L_- [ (\psi_+^1)' (\psi_+^2)' + L_+ \psi_+^1 \psi_+^2]}{(\rho^4- \ell^4 L_+ L_-)} \right) \partial_+ \nn
&+& \frac{\ell^2 \rho^2[ (\psi_+^1)' (\psi_+^2)' + L_+ \psi_+^1 \psi_+^2]}{(\rho^4- \ell^4 L_+ L_-)} \partial_-, 
\eea
in precise agreement with (\ref{Killing_vec}) and in line with our expectations. Once again, we stress the redundancy of the third order equation (\ref{3rd_order_eqn}); when the Killing vectors are expressed in terms of the underlying Killing spinors, both of which are locally valid, solutions to the second order Hill equation naturally emerge. As already highlighted in the introduction, here again we recognise that supersymmetry has played a key r\^ole turning a third order equation into a second order equation. This suppression of derivatives is a hallmark of supersymmetry. 

\section{Classification of geometries}
\label{sec:classification}
\setcounter{equation}{0} 
In this section we classify the Ba\~nados geometries according to the number of globally preserved supersymmetries.  From \eqref{KS} we see that a generic Killing spinor is not necessarily periodic or anti-periodic on angular coordinates $x^\pm$, therefore, it is not globally well-defined. As a result, the problem of finding periodic or anti-periodic Killing spinors reduces to finding periodic or anti-periodic solutions to the Hill equation \eqref{Hill}.

 It is useful to note that the Floquet index of solutions to the Hill equation, namely each $\psi_{\pm}$, is the same for all elements in a given Virasoro co-adjoint orbit \cite{Sheikh-Jabbari:2016unm}. In other words, all elements in a given orbit behave the same under $x^{\pm}\rightarrow x^\pm+2\pi$. Therefore, the amount of the global supersymmetry is an orbit invariant quantity. So, identifying the global supersymmetry of geometries corresponding to different representatives of Virasoro co-adjoint orbits gives a complete supersymmetry analysis of Ba\~nados geometries.

We recall that Virasoro co-adjoint orbits and the associated geometries were discussed in detail in the literature, and we refer interested readers to ref. \cite{Balog:1997zz, Sheikh-Jabbari:2016unm}. Here we briefly review the salient features.   {In each case we suppress indices on $x^{\pm}$, $\psi_{\pm}$ and $L_{\pm}$, and simply refer to them as $x$, $\psi$ and $L$, respectively, recalling from the previous section that they appear in separate Hill equations and Killing spinors. As a result, we can discuss them independently below.} The four monodromy classes are:
\begin{itemize}
\item \emph{Exceptional orbits}, with constant representative:
\be\label{En-psi}
L_n=-\frac{n^2}{4}\,,\quad \psi^1_n=\sqrt{\frac{2}{n}} \sin\frac{nx}{2}, \quad \psi_n^2 = \sqrt{\frac{2}{n}}\cos\frac{nx}{2},\qquad n\in\mathbb{Z}^+\,.
\ee
In this case $\psi^i$ \footnote{  {Recall there are two linearly independent solutions to the Hill equation.}} is a  periodic or an anti-periodic function of the $x$ coordinate,   {and likewise for all elements in this orbit. It is worth taking note of the fact that $n$ is a positive integer.}

\item \emph{Elliptic orbits}, with constant representative:
\be\label{Cnu-orbits}
L_\nu=-\frac{\nu^2}{4}\,,\quad \psi^1_\nu=\sqrt{\frac{2}{\nu}}\sin\frac{\nu x}{2}, \quad \psi_\nu^2 = \sqrt{\frac{2}{\nu}}\cos\frac{\nu x}{2},\qquad \nu\notin\mathbb{Z}^+\,,\ee
  {Here, in contrast to Exceptional orbits, $\psi^i$  is neither a periodic nor an anti-periodic function} under $x\rightarrow x+2\pi$. As a result, there is no overlap between Exceptional and Elliptic orbits; note $\nu$ is not a positive integer. 
\vskip 2mm

\item \emph{Hyperbolic orbits} can be further split into  cases with a functional dependence, 
\bea
L_{n,b} &=& b^2+\frac{b^2+4n^2}{2F_{n,b}}-\frac{3n^2}{4F_{n,b}^2}, \nn
\psi^1_{n,b} &=& \frac{e^{bx}}{\sqrt{F_{n,b}}}\sqrt{\frac{2}{n}}\left(\frac{b}{n} \cos\frac{n x}{2}+\sin\frac{nx}{2}\right), \nn
\psi^2_{n,b} &=&  \frac{e^{-bx}}{\sqrt{F_{n,b}}}\sqrt{\frac{2}{n}} \cos\frac{n x}{2},\,\qquad b\in\mathbb{R}^+\, \quad n\in\mathbb{N},
\eea
where we have defined 
\be 
F_{n,b}(x)=\left( \cos \frac{nx}{2} \right)^2+\left(\sin\frac{nx}{2}+\frac{2b}{n}\cos\frac{nx}{2}\right)^2, 
\ee
and {orbits with a} constant representative element 
\be
L_b=b^2\,,\quad \psi^1_b=\sqrt{\frac{1}{{2b}}}\; e^{bx},\quad \psi^2_b = \sqrt{ \frac{1}{{2b}}}\;e^{-bx},\,\qquad b\in\mathbb{R}^+.
\ee
Once again, $\psi^i$ is neither periodic nor anti-periodic.
\vskip 2mm

\item \emph{Parabolic orbits}, which can be further separated into  orbits with
\bea
\label{Parabolicorbits}
L_{n;\pm} &=&\frac{n^2}{2H_n}-\frac{3n^2(1\pm\frac{1}{2\pi})}{4H_n^2}, \\
\psi^1_n &=& \frac{1}{\sqrt{H_n}} \left(\pm\frac{x}{2\pi}\sin\frac{nx}{2}-\frac{2}{n}\cos\frac{nx}{2}\right), \quad \psi^2_n = \frac{1}{\sqrt{H_n}}\sin\frac{nx}{2}\,,\qquad n\in\mathbb{N}, \nonumber
\eea
where we have defined, 
\be
H_n(x)=1\pm\frac{1}{2\pi}\sin^2\frac{nx}{2}, 
\ee
or   {orbits with} 
\be\label{constantParabolic orbits}
L=0\,,\qquad \psi^1=\frac{x}{\sqrt{2\pi}}, \quad \psi^2 = {\sqrt{2\pi}}.
\ee

As one can explicitly see, in this case $\psi^2$ is either a periodic or anti-periodic function, while the $\psi^1$ is quasi-periodic. 

\end{itemize}
  {Having reviewed the Virasoro co-adjoint orbits, including a representative in each case, we are now in a position to classify supersymmetric geometries. To this end our strategy will be as follows.} 
Since the Killing spinors   {only depend on either $x^+$ or $x^-$}, by studying the periodicity behaviour of each sector separately, one can determine the number of   {preserved supersymmetries}. As coordinates $x^+$ and $x^-$ are considered periodic (with period $2\pi$), Killing spinors are globally well-defined if   {they} are periodic or anti-periodic functions of $x^\pm$.   {Therefore,} by counting the number of globally well-defined Killing spinors we can classify Ba\~nados geometries.   {We now enumerate the possibilities in turn, starting from the case with maximal supersymmetry, namely four supercharges.}

\subsection{Geometries with four supercharges}
When all four Killing spinors are periodic or anti-periodic in $x^+$ and $x^-$ \footnote{  {We will refer to these as left and right sectors, respectively.}}, we get a   {maximally supersymmetric geometry}.   {This happens} when both left and right sectors   {correspond to} Exceptional orbits \eqref{En-psi}. As discussed in \cite{Sheikh-Jabbari:2016unm}, when $n_+=n_-=N$ these geometries  are $N$-fold covers of AdS$_3$. For $n_+\neq n_-$ these geometries 
correspond to particles on $N$-fold covers of AdS$_3$, with $N=\textrm{lcm}(n_+,n_-)$. Note that while the representatives have constant character functions $L_\pm$, generic descendants may not have constant character functions. Global AdS$_3$ is the representative element   {with} $n_\pm=1$. The corresponding Killing spinors are anti-periodic in $x^\pm$ and for this reason, global AdS vacuum is also called NS-vacuum.

\subsection{Geometries with three supercharges}
Since the left and right sectors in the Ba\~nados geometries can be treated independently, we can have geometries with an odd number of global supersymmetries.  When one sector, e. g. the left sector, supports two global supercharges - $L_-$ is in an Exceptional orbit - while the other sector admits one supercharge, we get geometries with three global supersymmetries. 

From \eqref{Parabolicorbits} and \eqref{constantParabolic orbits} we find that for elements in Parabolic orbits, one solution to the Hill equation \eqref{Hill} is always periodic or anti-periodic, while the other one is not, and this sector admits only one global Killing spinor.  Therefore, if the geometry consists of an Exceptional orbit along with a Parabolic orbit, it admits three global Killing Spinors.

\subsection{Geometries with two supercharges}
There are two   {classes} of solutions with two global supercharges. This happens when each sector admits only one supercharge, i. e. they are in Parabolic orbits. The simplest example in this class is the null self-dual AdS$_3$ orbifold \cite{NSDO}, which is the near-horizon limit of massless BTZ \cite{MLBTZ}, where both left and right sectors correspond to the zeroth order Parabolic orbits, or $L_+=L_-=0$. Its descendants and all other geometries with both sectors in Parabolic orbits preserve two supercharges.

The other class corresponds to geometries where only one sector, for example the right sector, admits two global supercharges.   {This is the case where one sector is in an Exceptional orbit}, while the other sector does not preserve any supercharges. The left sector in this case is then either in a Hyperbolic or Elliptic orbit.   {This second class includes geometries corresponding to chiral particles, namely those with equal mass and spin \cite{Particles}, on an $N$-fold cover of AdS$_3$. }

\subsection{Geometries with one supercharge}
When one sector, e. g.  right sector, is in a Parabolic orbit, while the other sector is either in a Hyperbolic or Elliptic orbit, the corresponding geometries admit only one supercharge. The simplest example in this class is the extremal BTZ black hole,   {where} the right sector is in the zeroth Parabolic orbit with $L_+=0$, while the other sector is in the zeroth Hyperbolic orbit with $L_-$ a positive constant. Its near-horizon geometry, or self-dual orbifold \cite{SDO}, is another example in this class. The other example is the geometry corresponding to a chiral particle on AdS$_3$, where one sector belongs to the zeroth Parabolic orbit and the other sector is in an Elliptic orbit.  

\subsection{Non-supersymmetric geometries}
Finally when both left and right sectors are in  Hyperbolic or Elliptic orbits, the corresponding geometry does not admit any global supercharges. The simplest example in this class is the non-extremal BTZ black hole, where both sectors are in the zeroth Hyperbolic orbit, i. e. $L_\pm$ are positive constants. Particles on AdS$_3$, where both sectors are in an Elliptic orbit, provide a second illustrative example in this class. 

\section{Co-dimension two surfaces from supersymmetry}
\label{sec:codim2}
In this section, we will use supersymmetry to identify space-like co-dimension two surfaces in the Ba\~nados geometries. We will show that one can recover known minimal and extremal surfaces without resorting to finding space-like geodesics in the bulk. In principle, we will be able to apply supersymmetry to identify extremal surfaces in more complicated theories in 3D, for example gauged supergravities \cite{Deger:1999st, deWit:2003ja}, especially off-shell supergravities, where the geometry may not be asymptotically AdS$_3$ \cite{Rocek:1985bk, Nishino:1991sr, Kuzenko:2011rd, Kuzenko:2013uya, Alkac:2014hwa} \footnote{Supersymmetric solutions to these theories have been classified in \cite{Deger:2013yla, Deger:2016vrn} and include warped AdS$_3$ solutions analogous to the non-supersymmetric versions popularised in \cite{Anninos:2008fx}.}, not to mention higher-dimensional theories. 

Our motivation to pursue this approach stems largely from supersymmetric probe branes in the context of string theory, where kappa-symmetry \cite{Becker:1995kb, Bergshoeff:1997kr} (see also \cite{Sorokin:1999jx}) is exploited to find supersymmetric embeddings. However, there is one notable distinction; in the spacetimes we consider supersymmetry is only defined locally, yet it is enough to determine co-dimension two surfaces in the AdS$_3$ bulk. 

We consider a one-dimensional object with worldvolume parameter $s$. Pulling back the metric from the background 3D spacetime, we can define an induced metric, which is essentially the norm of a velocity vector $V$,  
\be
V^2  =  \frac{\dd x^{\mu}}{\dd s}  \frac{\dd x^{\nu}}{\dd s} g_{\mu \nu} = \dot{x}^{\mu} \dot{x}^{\nu} g_{\mu \nu}, 
\ee
where $x^{\mu}$, $\mu = 0, 1, 2$ denote AdS$_3$ coordinates and $g_{\mu \nu}$ is the 3D metric. Next, we introduce the dreibein, $e^{\mu} = E^{\mu}_{~\nu} \dd x^{\nu}$, and construct a projection condition: 
\be
\Gamma = \frac{1}{V} \dot{x}^{\mu} E^{\nu}_{~\mu} \gamma_{\nu}, 
\ee
where $\gamma_{\mu}$ denote our 3D constant gamma matrices (\ref{gamma_matrices}). Imposing the projection condition,  
\be
\label{kappa}
\Gamma \epsilon = \pm \epsilon, 
\ee
where $\epsilon$ solves the Killing spinor equation (\ref{KSE_ads}), we will find the equations for a curve $x^{\mu}(s)$ that is locally half-BPS. By construction $\Gamma^2 =1$, so the velocity vector $\dot{x}^{\mu}$ is guaranteed to be space-like, but in contrast to when the curve is a geodesic, \textit{a priori} $V$ is not a constant.  

  {Before proceeding to some familiar examples, we will consider what information may be extracted from the projection condition acting on the Killing spinors for the general class of Ba\~nados geometries. }Taking the same sign for projection conditions in \eqref{kappa} for two Killing spinors and using solution \eqref{KS} we get following equations
\bea
\frac{\ell \dot\rho}{\rho}+ \ell\left(\dot{x}^+-\frac{\ell^2L_-\dot{x}^-}{\rho^2}\right)\frac{\psi_+'}{\psi_+}=\pm V,\qquad \frac{\ell \dot\rho}{\rho}+ \ell\left(\dot{x}^--\frac{\ell^2L_+\dot{x}^+}{\rho^2}\right)\frac{\psi_-'}{\psi_-}=\mp V\,.
\eea

  {It is possible to find a solution to both} the equations (for plus sign in \eqref{kappa}),  
\be\label{kappa-solution}
\rho^2=\ell^2 \frac{\psi_+'}{\psi_+} \frac{\psi_-'}{\psi_-}\;.
\ee
It is useful to note that, in the above equation if we write $\psi_\pm$ in the form used in the Floquet Theorem, the above curve is the bifurcate Killing horizon of the Ba\~nados geometry, provided $\psi_\pm$ have real Floquet index \cite{Sheikh-Jabbari:2014nya}. Here we take general solutions of the Hill equation and \eqref{kappa-solution} includes the Killing horizon, but is not restricted to it. However, it is straightforward to check that, as we shall explain in the following examples, if we restrict to constant $\tau$ slices, where $\tau$ is an appropriate  time-like direction in the Ba\~nados geometries \cite{Sheikh-Jabbari:2016unm}, the resultant curve is a space-like geodesic. More concretely, the direction $\tau$ is defined to be a time-like Killing vector $\partial_{\tau}= \mathcal{K}^0_+ + \mathcal{K}^0_-$, where $\mathcal{K}^0_+$ is explicitly given in \eqref{PKV} and $\mathcal{K}^0_-$ is the vector with $+ \leftrightarrow -$. These vectors are related to the functions $K^0_{\pm}$ defined earlier in (\ref{vector_functions}).

To illustrate the workings of our projection condition, it is instructive to consider a simple example. We start with massless BTZ, where henceforth we set $\ell =1$ as it will not affect the analysis. Modulo a rewriting of the coordinates $x^{\pm}$, the metric takes form 
\be
\dd s^2 = r^2(- \dd t^2 + \dd x^2 ) + \frac{\dd r^2}{r^2}, 
\ee 
which is essentially the Poincar\'e metric modulo periodic boundary conditions on $x$, $t$ that break the superconformal supersymmetries. However, for the moment we relax global constraints and will just consider the local metric, where there is no distinction with Poincar\'e patch. We will in addition consider the simplification where we consider co-dimension two surfaces that do not depend on $t$, so that $\dot{t} = 0$. 

In Poincar\'e coordinates, the solutions to the Killing spinor equation (\ref{KSE_ads}) with both signs are,  
\bea
\label{epsilon_plus}
\epsilon_+ &=& \left( r^{-\frac{1}{2}} + r^{\frac{1}{2}} x \gamma_1 \right) \eta_{-} + r^{\frac{1}{2} } \eta_{+}, \nn
\label{epsilon_minus}
\epsilon_- &=& \left( r^{-\frac{1}{2}} -r^{\frac{1}{2}} x \gamma_1 \right) \tilde{\eta}_{+} + r^{\frac{1}{2} } \tilde{\eta}_{-}, 
\eea
where $\eta_{\pm}$ and $\tilde{\eta}_{\pm}$ are constant spinors satisfying $\gamma_r \eta_{\pm} = \pm \eta_{\pm}$ and $\gamma_r \tilde{\eta}_{\pm} = \pm \tilde{\eta}_{\pm}$. Since $t$ is a constant, we have absorbed it into $\eta_-$ and $\tilde{\eta}_{+}$, respectively. Otherwise, these are the expected form for the Killing spinors in Poincar\'e AdS$_3$ \cite{Lu:1998nu}. 

In our simplified setting, the projector may be expressed as 
\be
\Gamma = \frac{ {+} \dot{r} \gamma_r + r^2 \dot{x} \gamma_1 }{\sqrt{\dot{r}^2 + r^4 \dot{x}^2}}. 
\ee
Acting in turn on $\epsilon_{\pm}$, we arrive at two conditions: 
\bea
\label{massless_cond1}
\left[ - \frac{1}{r^3} \dot{r}  + x \dot{x} - \frac{1}{r^3} \sqrt{ r^4 \dot{x}^2 + \dot{r}^2 } \right] \eta_{{-}} &=& -\dot{x} \gamma_1 \eta_{{+}}, \\
\label{massless_cond2}
 \left[ \frac{1}{r^3} \dot{r} - x \dot{x}  - \frac{1}{r^3}  \sqrt{ r^4 \dot{x}^2 + \dot{r}^2 } \right] \tilde{\eta}_{{+}} &=& - \dot{x} \gamma_1 \tilde{\eta}_{{-}}, 
\eea
where we have opted for the positive sign in the projection condition (\ref{kappa}). We now recall that the spinors are non-zero constants \footnote{As discussed in the last section, global constraints determine if the constant spinors are zero or not.}, which allows us to introduce two constants above. Concretely, we rewrite the above equations as 
\bea
\label{massless_cond3} - \frac{1}{r^3} \dot{r}  + x \dot{x} - \frac{1}{r^3} \sqrt{ r^4 \dot{x}^2 + \dot{r}^2 } &=& - c \dot {x}, \\
\label{massless_cond4} \frac{1}{r^3} \dot{r} - x \dot{x}  - \frac{1}{r^3}  \sqrt{ r^4 \dot{x}^2 + \dot{r}^2 } &=& - \tilde{c} \dot{x},  
\eea
where we have introduced real constants, $c, \tilde{c}$, such that $\eta_{{-}}  = c \gamma_{1} \eta_{{+}}$, $\tilde{\eta}_{{+}} = \tilde{c} \gamma_1 \tilde{\eta}_{{-}}$. 

In arriving at (\ref{massless_cond3}) and (\ref{massless_cond4}), given (\ref{massless_cond1}) and (\ref{massless_cond2}), it is good to recall that $\eta_{\pm}$ and $\tilde{\eta}_{\pm}$ are both chiral. In particular focusing on $\eta_{\pm}$, we can write the spinors as 
\be
\eta_+ = \left( \begin{array}{c} \alpha \\ 0 \end{array} \right), \quad \eta_- = \left( \begin{array}{c} 0 \\  \beta \end{array} \right),
\ee
where $\alpha, \beta \in \mathbb{C}$ are constant. Since full supersymmetry is always present at a local level, $\alpha, \beta$ are non-zero. Plugging these explicit expressions for the spinors back into (\ref{massless_cond1}) we find: 
\be
\left[ - \frac{1}{r^3} \dot{r} + x \dot{x} - \frac{1}{r^3} \sqrt{r^4 \dot{x}^2 + \dot{r}^2} \right] {\beta} ={+} i \dot{x} {\alpha}.  
\ee
Taking either the real or imaginary part of the above equation, one gets (\ref{massless_cond3}). Similar logic applies in deducing (\ref{massless_cond4}). 

Subtracting the equations (\ref{massless_cond3}) and (\ref{massless_cond4}), we get 
\bea
\label{circle} -\frac{1}{r^3} \dot{r} + x \dot{x} &=& 0, 
\eea
where we have exploited shift-symmetry in {the isometry direction $x$} to remove a constant. Integrating (\ref{circle}) we arrive at the Ryu-Takayanagi embedding in Poincar\'e, 
\be
\frac{1}{r^2} + x^2 = h^2, 
\ee
where $h$ is an integration constant. We have recovered precisely the space-like geodesic for massless BTZ with constant $t$ (\ref{geo_massless_BTZ}), which we collect with other solutions to the geodesic equation in the appendix. Modulo shift symmetry in the isometry direction $x$, this is the most general solution that satisfies both (\ref{massless_cond1}) and (\ref{massless_cond2}). We can also confirm that the final form of the projection condition is consistent with supersymmetry: 
\be
\gamma_1 \eta_{{+}} = h \eta_{-}, \quad \gamma_1 \tilde{\eta}_{{-}} = h \tilde{\eta}_{{+}}. 
\ee
Since the projection condition acts on both the spinors, locally half the supersymmetry is preserved. 

As a further example, we consider minimal surfaces on the constant $t$ slice for static (non-rotating) BTZ, 
\be
\label{static_BTZ}
\dd s^2 = - (r^2 -m) \dd t^2 + \frac{\dd r^2}{(r^2-m)} + r^2 \dd x^2. 
\ee
Here, it is convenient to redefine $r = \sqrt{m} \cosh \rho$. Taking into account the sign choice in the Killing spinor equation, we find the solution 
\be
\epsilon_{\pm} = e^{ {\pm} \frac{\rho}{2} \gamma_r} e^{ {\pm} \frac{x \sqrt{m}}{2} \gamma_1} e^{ \frac{t \sqrt{m}}{2} \gamma_{r 0}} \eta_{\pm}, 
\ee
in terms of constant spinors $\eta_{\pm}$, where for constant $t$, we can drop the final exponential. In this case, our projection condition takes the form, 
\bea
\label{static_cond}
&& \biggl[ \dot{x}  \sqrt{m}  \cosh^2 \rho \gamma_1 + \gamma_r \left(  \dot{\rho} \cosh ( x \sqrt{m}) - \dot{x} \sqrt{m} \cosh \rho \sinh \rho \sinh (x \sqrt{m})\right) \\ && \pm \gamma_{r1} \left( \dot{\rho} \sinh ( x \sqrt{m}) - \dot{x} \sqrt{m} \cosh \rho \sinh \rho \cosh (x \sqrt{m})\right)  \biggr] \eta_{\pm}=  \sqrt{\dot{x}^2 m \cosh^2 \rho + \dot{\rho}^2} \, \eta_{\pm}.  
\nonumber
\eea
It is worth noting that we again have two equations, but now there is no chirality condition imposed on $\eta_{\pm}$. The above equations have the structure,
\be
[ A \gamma_1 + B \gamma_r \pm C \gamma_{r1} ] \eta_{\pm} = \eta_{\pm}. 
\ee
Using the explicit gamma matrices (\ref{gamma_matrices}) and constant spinors, 
\be
\eta_{\pm} = \left( \begin{array}{c} \alpha_{\pm} \\ \beta_{\pm} \end{array} \right), \quad \alpha_{\pm}, \beta_{\pm} \in \mathbb{C}, 
\ee
one can rewrite the equations as 
\be
i ( A \pm C) \frac{\beta_{\pm}}{\alpha_{\pm}} = (1-B). 
\ee
Once again, we use the fact that $\alpha_{\pm}, \beta_{\pm}$ are non-zero, before subtracting to infer that $C = \kappa A$, with constant $\kappa$, or more explicitly
\bea
\dot{\rho}  \sinh ( x \sqrt{m} ) - \dot{x} \sqrt{m} \cosh \rho \sinh \rho \cosh ( x \sqrt{m})  =  \kappa \,  \dot{x} \sqrt{m} \cosh^2 \rho, 
\eea
where $\kappa$ is a constant. Integrating this equation and absorbing the integration constant in a shift in the $x$-direction, we find
\be
\frac{\sqrt{r^2-m}}{r} =  \frac{\sqrt{r_*^2-m}}{r_*} \cosh ( x\sqrt{m} ). 
\ee
Note, here we have set the remaining constant by setting $r = r_*$ to be the value of $r$ when $x = 0$.  This is the most general solution to the two conditions in (\ref{static_cond}). Up to the redefintion $r_* = L$, this is the expression for the relation implied by the constant $t$ geodesic (\ref{geo_static_BTZ}). It also agrees with (5.32) of ref. \cite{Hubeny:2007xt}, when one realises that (5.32) may be further simplified and recast as 
\be
x = \frac{1}{\sqrt{m}} \log \left( \frac{r \sqrt{r_*^2-m}}{r_* \sqrt{r^2-m}- \sqrt{m} \sqrt{r^2 -r_*^2}}\right). 
\ee

Finally, we record the final projection condition on the constant spinor
\be
\label{projection_BTZ}
\left( \frac{r_*}{\sqrt{m}} \gamma_x - \sqrt{\frac{r_*^2}{m}-1} \gamma_{rx} \right) \epsilon_0 = \epsilon_0, 
\ee
which clearly shows that supersymmetry is preserved, at least at the local level. 

Before proceeding to rotating spacetimes, we pause for one final comment. Given Killing spinors, one can construct a vector, or one-form, $\mathcal{K}^{\pm}_{\mu}$ (\ref{vec_bilinear}), which is either time-like or null. This presents us with a natural one-form as a potential candidate for a calibration. However, for space-like configurations we consider, it is easy to show that $\mathcal{K}^{\pm}_{\mu}$ is trivially zero. This is easily understood as the outcome of reconciling supersymmetry, which demands time-like or null, and the space-like condition we impose by hand. In fact, one can use the vanishing of $\mathcal{K}^{\pm}_{\mu}$ to determine constant $\tau$ geodesics in AdS$_3$. 

\subsection{Extremal co-dimension two surfaces} 

As we have seen, our projection condition agrees with known results for static spacetimes, where the original prescription of Ryu-Takayanagi \cite{Ryu:2006bv} allows one to determine the minimal surfaces. We should now show the utility of this method in a time-dependent configuration, which means we should be able to apply the same methodology to determine the extremal surfaces for the rotating BTZ black hole. In contrast to the static case, where supersymmetric surfaces in Riemannian manifolds can be related to calibrated cycles \cite{Harvey:1982xk}, which are minimal surfaces in their homology class, there is no generalisation to extremal surfaces in pseudo-Riemannian manifolds. 

To apply our projection condition, we require a knowledge of the local Killing spinors in the rotating BTZ background. Recall again that supersymmetry is broken globally, but AdS$_3$ is quirky and we can still find local solutions. To find the Killing spinors, it is best to realise that the rotating BTZ solution may be generated via a boost starting from static BTZ (\ref{static_BTZ}). Concretely, one considers the transformation, 
\bea
\left( \begin{array}{c} \dd t \\ \dd x \end{array} \right) \rightarrow \left( \begin{array}{cc}  \cosh \gamma & \sinh \gamma \\ \sinh \gamma & \cosh \gamma \end{array} \right)  \left( \begin{array}{c} \dd t \\ \dd x \end{array} \right), 
\eea
while redefining: 
\bea
\tilde{r}^2 &=& r^2 + m \sinh^2 \gamma, \quad \tilde{r}_+ = \sqrt{m} \cosh \gamma, \quad \tilde{r}_- = \sqrt{m} \sinh \gamma, 
\eea
and dropping tildes. The resulting metric is (\ref{BTZ}). 

Using the fact that the rotating BTZ metric is simply the boosted static solution, it is easy to infer that the Killing spinors change due to a frame rotation. As a result of the frame rotation, the Killling spinor becomes \footnote{We use the dreibein, $e^{t} = \frac{\sqrt{(r^2-r_+^2)(r^2-r_-^2)}}{r} \dd t, ~e^{r} = \frac{r}{\sqrt{(r^2-r_+^2)(r^2-r_-^2)}} \dd r, ~e^{x} = r ( \dd x + \frac{r_+ r_-}{r^2} \dd t)$.}
\be
\label{KS_BTZ}
\epsilon_{\pm} = e^{ \frac{\alpha_{\pm}}{2} \gamma_r} e^{\frac{1}{2}(X_{+} \pm X_-) \gamma_1} \epsilon_0, 
\ee
where we have used $\gamma_{0 1 r} = 1$ and defined: 
\bea
X_{\pm} &=& x r_{\pm} + t r_{\mp}, \nn
\cosh \alpha_{\pm} &=&  \frac{r_+ r_- \pm r^2}{r(r_+ \pm r_-)} ~~ \Rightarrow ~~ \sinh \alpha_{\pm} = \frac{\sqrt{(r^2-r_+^2)(r^2-r_-^2)}}{r (r_+ \pm r_-)}
\eea

Upon expansion of the projection condition we encounter the following condition:
\bea
\label{projection}
\frac{1}{V} \biggl[  A_{\pm} \gamma_1 + B_{\pm} \gamma_r + C_{\pm} \gamma_{1r} \biggr] \epsilon_0 = \pm \epsilon_0, 
\eea 
where we have defined: 
\bea
A_{\pm} &=& \frac{1}{(r_+^2 - r_-^2)}  \left[ \pm \dot{X}_+ (r^2-r_-^2) - \dot{X}_- (r^2 - r_+^2) \right], \nn
B_{\pm} &=&  \frac{\cosh (X_+ \pm X_-) \, r \dot{r}}{\sqrt{(r^2-r_+^2)(r^2-r_-^2)}} - \sinh (X_+ \pm X_-) \frac{\sqrt{(r^2-r_+^2)(r^2-r_-^2)}}{(r_+^2-r_-^2)} (\dot{X}_+ \mp \dot{X}_-), \nn
C_{\pm} &=& \cosh (X_+ \pm X_-) \frac{\sqrt{(r^2-r_+^2)(r^2-r_-^2)}}{(r_+^2-r_-^2)} (\dot{X}_+ \mp \dot{X}_-) - \frac{ \sinh (X_+ \pm X_-) \, r \dot{r}}{\sqrt{(r^2-r_+^2)(r^2-r_-^2)}}, \nn
V^2 &=&  \frac{-\dot{X}_-^2 (r^2 -r_+^2) + \dot{X}^2_+ (r^2 - r_-^2)}{(r_+^2-r_-^2)}  + \frac{r^2  \dot{r}^2} {(r^2-r_+^2)(r^2-r_-^2)}. 
\eea

Given the complexity of the two projection conditions (\ref{projection}), we will adopt a slightly different strategy here to solve them. We note that (\ref{projection}) is expected to preserve supersymmetry for both signs on the RHS of the equation. Earlier, we simply opted for the positive sign, but it is good to recall that the final projection condition we obtain should work with either choice of sign. This can be clearly seen from (\ref{projection_BTZ}), where we can flip the sign on the RHS and get an equally valid projection condition - just a different relation between spinor components.

Using both signs, we can deduce that the coefficients of the gamma matrices on the LHS must be constant. To better understand this claim, let us rewrite (\ref{projection}) as 
\be
\label{ABC_constant}
\biggl[  A \gamma_1 + B \gamma_r + C \gamma_{1r} \biggr] \epsilon_0 = \pm \epsilon_0, 
\ee
where we have dropped subscripts and absorbed $V$. Given there is a constant $\eta_0$ for both signs,  we can infer 
\be
i (A-C)  = \kappa_1 (1 - B), \quad i (A - C )  =  \kappa_2(-1 - B), 
\ee
where $\kappa_i$ denote the ratio between constant spinors in $\epsilon_0$ for the two choices of sign. If $\kappa_1 = - \kappa_2$, we immediately have $B=0$, but irrespective of the choice of $\kappa_1, \kappa_2$, provided they are non-zero, we infer $B$ is a constant through subtraction. This further implies that $A-C$ is a constant. Finally, from the square of (\ref{ABC_constant}), we have $A^2 + B^2 - C^2 = 1$, which implies that both $A$ and $C$ are constant. This completes our demonstration that $A_{\pm}/V$, etc in (\ref{projection}) must be constants. 

Armed with this information, we can deduce that $B_{\pm}$ and $C_{\pm}$ have to related up to constants, which gives us two equations. Shifting $X_+$ and $X_-$ accordingly, one can eliminate the $\dot{r}$ term to solve $X_-$ in terms of $X_+$ 
\bea
\label{XpXm}
\tanh( X_+ + X_-) (\dot{X}_+ - \dot{X}_-) &=& \tanh (X_+ - X_-) (\dot{X}_+ + \dot{X}_-) ~~ \Rightarrow  \nn
\tanh X_- &=&  \kappa  \tanh X_+, 
\eea
where $\kappa$ is an integration constant. 

We next eliminate $X_-$ to find the following equation: 
\bea
\frac{(r_+^2 -r_-^2) r \dot{r}}{(r^2-r_+^2)(r^2-r_-^2)} = \frac{(1-\kappa^2) \cosh X_+ \sinh X_+ \dot{X}_+}{(\cosh^2 X_+ - \kappa^2 \sinh^2 X_+)}, 
\eea
which may be easily integrated to find
\be
\label{rotating_BTZ_eqn}
\sqrt{\frac{(r^2-r_+^2)}{(r^2-r_-^2)}} = \lambda \frac{\cosh X_+}{\cosh X_-}, 
\ee
where $\lambda$ is another integration constant. Note that in writing the final expression we have used (\ref{XpXm}) to reintroduce $X_-$.  This is the most general solution to (\ref{projection}). 

As a check that all steps have been performed correctly, we can reinsert (\ref{rotating_BTZ_eqn}) back into the projection condition (\ref{projection}) to recover 
\be
\frac{1}{\sqrt{(1-\lambda^2)(1-\kappa^2 \lambda^2)} } \left[  \pm (1 \mp \kappa \lambda^2) \gamma_1 + (1 \mp \kappa) \lambda \gamma_{1r} \right] \epsilon_0 = \pm \epsilon_0.  
\ee
We observe that all the coefficients of the gamma matrices are indeed constant, so this is a good projection condition. We can now also record the velocity vector,  
\be
 V  = \frac{(r^2-r_-^2)}{(r_+^2 -r_-^2)} \dot{X}_+ \sqrt{(1-\lambda^2)(1-\kappa^2 \lambda^2)}. 
\ee
It is worth noting that the velocity vector is not a constant, so we have yet to recover the space-like geodesic. On the contrary, using the expressions in the appendix, it is easy to show there is a space-like geodesic with two constants of motion, which satisfies (\ref{XpXm}) and (\ref{rotating_BTZ_eqn}). At this juncture, we could, as suggested earlier, impose constant $\tau$, where $\tau$ is the 
time-like direction defined in the Ba\~nados geometries, but this imposes $X_-$ constant. 

Instead, by setting $V= 1$, we find that the integrated geodesic equations, namely (\ref{dott}), (\ref{dotx}) and (\ref{dotr}) are recovered, provided: 
\be
E = \frac{r_+ \, \kappa \lambda^2 -r_-}{\sqrt{(1-\lambda^2)(1-\kappa^2 \lambda^2)}}, \quad L = \frac{r_+ - r_- \, \kappa \lambda^2}{\sqrt{(1-\lambda^2)(1-\kappa^2 \lambda^2)}}. 
\ee

Solving for $\kappa, \lambda$ one finds two solutions, which are simply related through the interchange $r_+ \leftrightarrow r_-$. More precisely, we have 
\be
\lambda = \sqrt{\frac{(\alpha -2 r_+^2 +\gamma)}{(\alpha -2 r_-^2 + \gamma)}}, \quad \kappa = \frac{(\alpha -2 r_+^2 -\gamma) (E r_- + L r_+)}{(\alpha -2 r_-^2 -\gamma)(E r_+ + L r_-)}, 
\ee
where the additional constants $\alpha, \gamma$ are defined in the appendix. 

So to summarise and close this section, for rotating BTZ, imposing supersymmetry and $V=1$, we can recover the solution to the geodesic equation (\ref{BTZ_geo_solution}) quoted in the appendix.

\section{Conclusions \& Outlook}
\label{sec:conclusion} 
In the first part of this work we extended the classification of AdS$_3$ solutions preserving global supersymmetry to the Ba\~nados class of geometries, thereby completing work initiated in ref. \cite{Coussaert:1993jp}. Locally, we provided the solution to the Killing spinor equation for the Ba\~nados class and showed that Killing spinors are in one-to-one correspondence with solutions to the Hill differential equation. From the Killing spinors, we illustrated how one reconstructs locally the $SL(2, \mathbb{R}) \times SL(2, \mathbb{R})$ isometries of the Ba\~nados geometries. Using periodicity properties of the Killing spinors, we provided examples of Ba\~nados geometries preserving zero, one, two, three and four supersymmetries.

It should be possible to repeat the asymptotic symmetry group analysis, \`a la Brown-Henneaux \cite{Brown:1986nw}, for our supersymmetric theory, namely with super-diffeomorphisms.  It is expected to realise a super-Virasoro algebra. In the same way that Ba\~nados geometries can be classified with Virasoro co-adjoint orbits, we expect supersymmetric Ba\~nados geometries may be classified using super-Virasoro orbits.

In the second part of this paper, we introduced a projection condition, which allowed us to isolate half-BPS space-like co-dimension two surfaces in the bulk AdS$_3$ spacetime. More precisely, we demonstrated that for constant $\tau$, where $\tau$ is a time-like direction defined in the Ban\~ados geometries \cite{Sheikh-Jabbari:2016unm} \footnote{See earlier comments below (4.5) of the text.}, one recovers space-like geodesics in the general class of Ba\~nados geometries, including massless, static and rotating BTZ. For rotating BTZ, we observed that a more general class of space-like geodesics could be recovered by instead imposing that the velocity vector of the supersymmetric curve has unit norm. With these assumptions, we have shown that supersymmetry recovers the known HRT extremal surfaces in 3D without exploiting i) equivalence of geometries to Poincar\'e AdS$_3$, or ii) calculating space-like geodesics. 

We have swept two technical details under the rug in this work. Firstly, we should identify how our projection condition is related to the geodesic equation. It is clear that the geodesic equation is not naively the ``square" of the projection condition, since we do not recover the geodesics uniquely, unless we impose $\tau$ constant. Secondly, our approach is motivated by calibrations \cite{Harvey:1982xk}, yet in pseudo-Riemannian manifolds, such a concept does not apply. Potentially by analytically continuing to Euclidean signature, or by restricting to a constant $\tau$ slice, one can elucidate the relation between our projection condition and a calibrated cycle. It is likely these two issues are related. 

There are a number of open directions. Firstly, it would be interesting to look for extremal surfaces in 3D gauged supergravities, where additional scalar matter is present. These theories are embeddable in string/M-theory \cite{Detournay:2012dz, Karndumri:2013dca, Karndumri:2015sia}, so one has greater control over the dual CFT. Extending the analysis to off-shell supergravity in 3D, a host of supersymmetric warped AdS$_3$ vacua exist \cite{Deger:2013yla, Deger:2016vrn}, so one should be able to identify extremal surfaces in this setting. On that note, a candidate extremal surface for warped AdS$_3$ was recently identified \cite{Song:2016pwx}, where it was noted that the curve was not necessarily a geodesic. It remains an exercise to derive this curve from supersymmetry. 

Finally, it would be of considerable interest to extend our work to higher dimensions, where the difficulties determining extremal surfaces, at least analytically, using the HRT prescription \cite{Hubeny:2007xt} are well documented. It is conceivable that by combining HRT with supersymmetry, one may be in a position to identify higher-dimensional extremal surfaces that do not correspond to simple strips and balls on the AdS boundary.

\section*{Acknowledgements} 
We are grateful to J. Gauntlett, D. Giataganas and C. Y. Park for discussions. We acknowledge M. M. Sheikh-Jabbari for the initial discussion, along with his insights, which helped seed this project. We thank D. Giataganas, C. Y. Park, M. M. Sheikh-Jabbari and T. Takayanagi for comments on the final draft. E. \'O C is supported by the Marie Curie Fellowship PIOF-2012- 328625 T-DUALITIES.

\section*{Data Management}
No additional research data beyond the data presented and cited in this work are needed to validate the research findings in this work.

\appendix 

\section{Floquet Theorem}
In this appendix, we review the Floquet theorem following \cite{Hill}. Consider $L(x)$, a complex function of a real variable $x$, which is piecewise continuous and periodic, 
\be
L (x + 2\pi) = L(x), \quad \forall x \in \mathbb{R}. 
\ee
If $L(x)$ has these properties, then the Hill differential equation 
\be
\label{diff_eqn}
\psi'' - L(x) \psi = 0, 
\ee
has two continuously differentiable solutions $\psi_1(x), \psi_2(x)$, which are uniquely determined by the conditions: 
\be
\psi_1(0) = 1, \quad \psi_1'(0) = 0, \quad \psi_2(0) =0, \quad \psi_2'(0) =1. 
\ee
Associated to (\ref{diff_eqn}), we can define a characteristic equation
\be
\rho^2 - [ \psi_1(2\pi) + \psi_2'(2\pi) ] \rho +1 = 0, 
\ee
and a characteristic exponent  $\alpha$ (Floquet index), such that 
\be
e^{i \alpha \pi} = \rho_1, \quad e^{- i \alpha \pi} = \rho_2, 
\ee
where $\rho_1, \rho_2$ are the roots of the characteristic equation. 

\noindent
\textbf{Floquet theorem.} \\
{If the roots $\rho_i$ of the characteristic equation are different from each other, then Hill equation has two linearly independent solutions} 
\be
\psi_1(x) = e^{i \alpha x} p_1 (x), \quad \psi_2(x) = e^{- i \alpha x} p_2(x), 
\ee
where $p_i(x)$ are periodic with period $2\pi$. If $\rho_1 = \rho_2$, then equation (\ref{diff_eqn}) has a nontrivial solution, which is periodic with period $\pi$ ($\rho_1 = \rho_2 = 1$) or $ 4 \pi$ (when $\rho_1 = \rho_2 = -1$). Let $p(x)$ denote such a periodic solution and let $\psi(x)$ be another solution linearly independent of $p(x)$. Then, 
\be
\psi(x + 2\pi) = \rho_1 \psi(x) + \theta p (x), \quad \theta ~~\textrm{constant}, 
\ee
and $\theta = 0$ is equivalent to 
\be
\psi_1(2\pi) + \psi_2'(2\pi) = \pm 2, \quad \psi_2(2\pi) = 0, \quad \psi_1'(2\pi) = 0. 
\ee

\section{Space-like geodesics for BTZ black hole}
In this section, we record the solution to the geodesic equation
\be
\frac{\dd ^2 x^{\mu}}{\dd s^2} + \Gamma^{\mu}_{\rho \sigma} \frac{\dd x^{\rho}}{\dd s} \frac{\dd x^{\sigma}}{\dd s} = 0, 
\ee
with affine parameter $s$ for the BTZ black hole (\ref{BTZ}) with unit radius $\ell =1$. Null and time-like geodesics can be found in ref. \cite{Cruz:1994ir} and here we focus on space-like geodesics. Once we have a general expression, we will spell out the simplifications for massless and static BTZ. 

To solve the geodesic equation, we will make use of the fact that $\partial_t$ and $\partial_{\phi}$ are Killing vectors, so that 
\be
E = - g_{t \mu} u^{\mu} , \quad L =  g_{\phi \mu} u^{\mu}, 
\ee
will be constants of motion along the geodesic for a given velocity vector $ u^{\mu} = \frac{\dd x^{\mu}}{\dd s}$. 
Inverting these expressions, we can determine $ \dot{t} = \frac{\dd t}{\dd s}$, $\dot{\phi} = \frac{\dd \phi}{\dd s}$, in terms of these constants: 
\bea
\label{dott} \dot{t} &=& \frac{r^2 E + r_+ r_- L}{(r^2-r_+^2)(r^2-r_-^2)} = \frac{1}{r_+^2-r_-^2} \left( \frac{r_+ (E r_+ + L r_-)}{(r^2-r_+^2)} - \frac{r_- (E r_- + L r_+ )}{(r^2-r_-^2)} \right), \\
\label{dotx} \dot{x} &=&  \frac{(r^2 - (r_+^2+r_-^2)) L - r_+ r_- E}{(r^2-r_+^2)(r^2-r_-^2)} = \frac{1}{r_+^2-r_-^2} \left( \frac{r_+ (E r_- + L r_+)}{(r^2-r_-^2)} - \frac{r_- ( E r_+ +L r_-  )}{(r^2-r_+^2)} \right). \nonumber
\eea
Next, we impose that the geodesic is space-like, $ u_{\mu} u^{\mu} = 1$, which allows us to determine $\dot{r}$ also in terms of $E$ and $L$, 
\be
\label{dotr} r^2 \dot{r}^2 = (E^2 -L^2) r^2 + 2 E L r_+ r_- + L^2 (r_+^2 +r_-^2) + (r^2-r_+^2) (r^2 - r_-^2). 
\ee
We can then integrate the three equations, in the process absorbing the three integration constants into shifts in the affine parameter $s$ and coordinates $t, x$. The final solution to the geodesic equation may be neatly expressed as 
\bea
\label{BTZ_geo_solution}
r^2 &=& \frac{1}{2} \left[ \gamma \cosh 2 s + \alpha \right], \nn
\tanh (r_+ x + r_- t ) &=& \frac{(L^2-E^2+r_+^2-r_-^2-\gamma)}{2 (L r_+ + E r_-)} \tanh s, \nn
\tanh (r_- x + r_+ t ) &=& \frac{(L^2-E^2-r_+^2+r_-^2-\gamma)}{2 (L r_- + E r_+)} \tanh s, 
\eea
where we have further defined,  
\bea
\label{alpha_gamma}
\alpha &=& (r_+^2 + r_-^2) + L^2 -E^2, \quad \beta = 2 E L r_+ r_- + L^2 (r_+^2 + r_-^2) + r_+^2 r_-^2, \nn
\gamma &=& \sqrt{\alpha^2 - 4 \beta}. 
\eea
The following relation is implied: 
\bea
\label{geo_rotating_BTZ}
\frac{\cosh X_+}{\cosh X_-} = \sqrt{\frac{\alpha -2 r_-^2 + \gamma}{\alpha -2 r_+^2 + \gamma}} \sqrt{\frac{(r^2-r_+^2)}{(r^2-r_-^2)}}. 
\eea

From here we can recover the geodesic for static BTZ by setting $r_- = 0, r_+ = \sqrt{m}$. Moreover, if we are interested in geodesics that are independent of $t$, we can truncate the above equations by setting $E = 0$. With $t$ constant, the above expressions simplify, 
\bea
r^2 &=& \frac{1}{2} \left[ (L^2 -m) \cosh 2 s + (L^2+m) \right], \quad
\tanh ( x \sqrt{m}) = \frac{\sqrt{m}}{L} \tanh  s, 
\eea
which implies the following relation: 
\bea
\label{geo_static_BTZ} \frac{\sqrt{r^2-m}}{r} = \frac{\sqrt{L^2-m}}{L} \cosh (x \sqrt{m}). 
\eea
One final simplification follows in the massless case once we set $m =0$ with $t$ constant: 
\bea
\label{geo_massless_BTZ}
r = L^2 \cosh^2 s, \quad x = \frac{1}{L} \tanh s \quad \Rightarrow \quad \frac{1}{r^2} + x^2 = \frac{1}{L^2}. 
\eea 

{}

\end{document}